\documentclass[journal]{IEEEtran}
\usepackage{amsmath,amsfonts}
\usepackage{algorithmic}
\usepackage{array}
\usepackage[caption=false,font=normalsize,labelfont=sf,textfont=sf]{subfig}
\usepackage{textcomp}
\usepackage{stfloats}
\usepackage{url}
\usepackage{verbatim}
\usepackage{graphicx}
\usepackage{cite}
\usepackage{hyperref}

\hypersetup{
    colorlinks=true,
    linkcolor=blue,
    filecolor=magenta,      
    urlcolor=cyan,
    pdftitle={Online Feature Decoupling Framework for High Fidelity Ultrasound Video Synthesis},
    pdfpagemode=FullScreen,
    }
\usepackage{booktabs} 
\usepackage{tabularx}
\usepackage{multirow} 

\newcommand{\ourmodel}{OnUVS}

\def\BibTeX{{\rm B\kern-.05em{\sc i\kern-.025em b}\kern-.08em
    T\kern-.1667em\lower.7ex\hbox{E}\kern-.125emX}}
\usepackage{balance}
\begin{document}

\title{\ourmodel: Online Feature Decoupling Framework for High-Fidelity Ultrasound Video Synthesis}
\author{Han Zhou, Dong Ni, Ao Chang, Xinrui Zhou, Rusi Chen, Yanlin Chen, Lian Liu, Jiamin Liang, Yuhao Huang, Tong Han, Zhe Liu, Deng-Ping Fan, Xin Yang 
\thanks{Han Zhou, Dong Ni, Ao Chang, Xinrui Zhou, Rusi Chen, Yanlin Chen, Lian Liu, Jiamin Liang, Yuhao Huang, Tong Han, Zhe Liu, Xin Yang are with National-Regional Key Technology Engineering Laboratory for Medical Ultrasound, School of Biomedical Engineering, Health Science Center, Shenzhen University, China}
\thanks{Lian Liu is with National-Regional Key Technology Engineering Laboratory for Medical Ultrasound, School of Biomedical Engineering, Health Science Center, Shenzhen University, and Shenzhen RayShape Medical Technology Co., Ltd, China.}
\thanks{Deng-Ping Fan is with Computer Vision Lab (CVL), ETH Zurich, Zurich, Switzerl.}
\thanks{Corresponding author: Xin Yang (xinyang@szu.edu.cn).}
\thanks{Han Zhou and Dong Ni contributed equally to this work.}}


\maketitle

\begin{abstract}
Ultrasound (US) imaging is indispensable in clinical practice. To diagnose certain diseases, sonographers must observe corresponding dynamic anatomic structures to gather comprehensive information. However, the limited availability of specific US video cases causes teaching difficulties in identifying corresponding diseases, which potentially impacts the detection rate of such cases.
The synthesis of US videos may represent a promising solution to this issue. Nevertheless, it is challenging to accurately animate the intricate motion of dynamic anatomic structures while preserving image fidelity.
To address this, we present a novel online feature-decoupling framework called \ourmodel~for high-fidelity US video synthesis. Our highlights can be summarized by four aspects.
First, we introduced anatomic information into keypoint learning through a weakly-supervised training strategy, resulting in improved preservation of anatomical integrity and motion while minimizing the labeling burden.
Second, to better preserve the integrity and textural information of US images, we implemented a dual-decoder that decouples the content and textural features in the generator.
Third, we adopted a multiple-feature discriminator to extract a comprehensive range of visual cues, thereby enhancing the sharpness and fine details of the generated videos.
Fourth, we constrained the motion trajectories of keypoints during online learning to enhance the fluidity of generated videos. Our validation and user studies on in-house echocardiographic and pelvic floor US videos showed that \ourmodel~synthesizes US videos with high fidelity.
\end{abstract}


\section{Introduction}
\begin{figure*}[!t]
\centering
\includegraphics[width=0.92\linewidth]{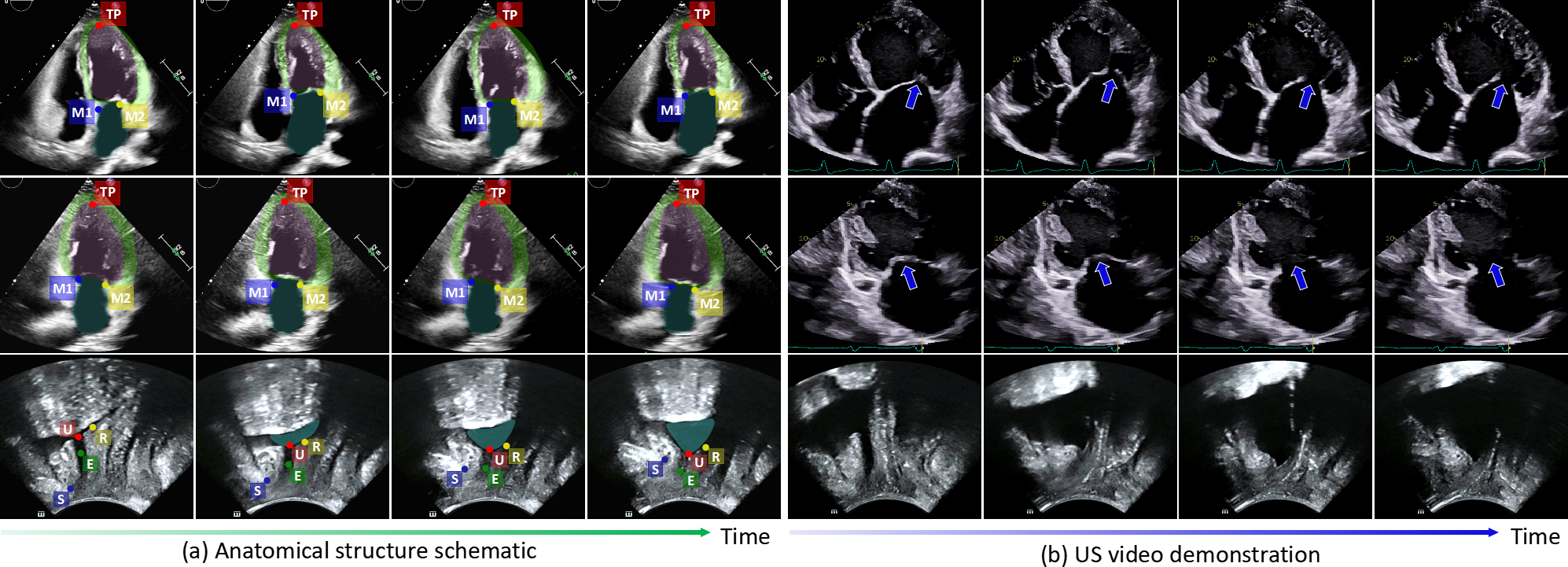}
\caption{Transthoracic echocardiography and pelvic floor US videos. From top to bottom, apical four-chamber (A4C) heart echocardiography, apical two-chamber (A2C) heart echocardiography, and pelvic floor US. TP denotes the heart tip point. M1 and M2 represent mitral valve annulus. S, U, E, and R are the pubic bone, urethral meatus, midpoint of the urethra, and posterior bladder wall incision point, respectively.}
\label{fig_us_image_intro}
\end{figure*}
\IEEEPARstart{U}{ltrasound} (US) imaging is a non-invasive, radiation-free, and real-time approach to visualize the morphology and movements of internal organs. As an essential medical imaging modality, it has extensive applications in clinical practice, including transthoracic echocardiography (TTE) and pelvic floor US.
TTE (first two rows in Fig.~\ref{fig_us_image_intro}) illustrates the morphology of the heart, and the periodic activity of its walls, ventricles, and valves. TTE is the first-line imaging tool for the diagnosis, evaluation, and management of suspected or confirmed heart disease \cite{corbett2022practical}.
Pelvic floor US (bottom row in Fig.~\ref{fig_us_image_intro}) is used to diagnose pelvic floor dysfunction, which is estimated to affect as many as 50 percent of women over the age of 50 \cite{bahrami2021pelvic}. Using US imaging, the static and dynamic structures of the pelvic floor in the resting, retracted, and Valsalva states can be observed. This enables the timely detection of morphological changes in said structures prior to the development of clinical symptoms, thereby preventing further progression of such changes.  
Pelvic floor US has become the preferred imaging modality for diagnosing pelvic floor dysfunction disorders \cite{testtttt}.\par
TTE and pelvic floor US imaging are examples of examinations that rely on dynamic videos to gather essential anatomic information for accurate clinical diagnoses. Because US video-based diagnoses require the observation of dynamic changes in anatomic structures, an abundant supply of such videos is necessary to train sonographers in diagnostic tasks. Unfortunately, low morbidity rates and limitations on clinical resources have resulted in a scarcity of US videos for specific cases. This scarcity complicates the training of novice sonographers for the detection of these cases, potentially influencing the overall detection rate of such cases. The generation of synthetic US videos has emerged as a promising solution to this problem by facilitating the training of sonographers \cite{bradley2020determining}.\par
A relatively early US video synthesis was performed by \cite{stoitsis2008simulating} for the dynamic carotid artery wall. Specifically, the anatomy of the carotid artery was modeled by generating a series of scatter maps representing the densities and sound velocities of various arterial tissues. Subsequently, US images were synthesized using FIELD II \cite{field2}, and the generation of a 29-frame sequence required approximately 14 h. \cite{alessandrini2012simulation} used real echocardiographic sequences as templates for the simulation of realistic cardiac US sequences. \cite{solomou2016ultrasound} applied motion in the axial and radial directions of consecutive video frames to synthesize carotid artery US videos. \cite{evain2022motion} used a physical simulator to synthesize echocardiogram US sequences. \par
The conventional approach to US video synthesis through physical modeling often fails to achieve sufficient realism and is resource- and time-intensive. Note that the aforementioned methods generate US videos frame-by-frame. A similar task, namely that of US image synthesis, is frequently performed via deep-learning techniques, which may provide inspiration for US video synthesis. \cite{peng2019real} inferred anatomic structures from magnetic resonance imaging data and synthesized US images using generative adversarial networks (GAN) \cite{goodfellow2020generative}. \cite{bargsten2020specklegan} synthesized intravascular US images by generating medical images exhibiting speckle noise with a GAN. \cite{che2021realisticliver} synthesized diseased and healthy liver US images. \cite{kumar2021empiricalUSS} synthesized US images to represent the variability of disease symptoms. \par
\begin{figure*}[!t]
    \centering
    \includegraphics[width=0.92\linewidth]{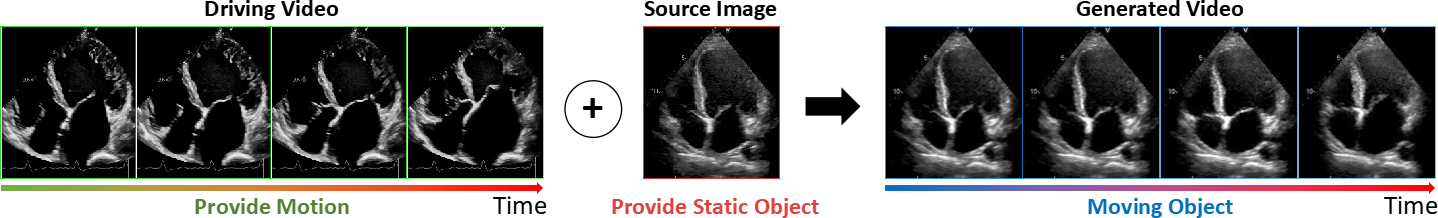}
    \caption{Generation of US video by transferring motion from a driving video to a static source image.}
    \label{fig: method intro}
\end{figure*}
Although US image synthesis has achieved remarkable results, significant disparities remain between it and the syntheses of US videos, particularly when considering the consistency of the highly dynamic motion of anatomic structures across frames. Consequently, these methods cannot be applied directly to synthesize US videos.
Meanwhile, in the field of natural images, there is growing attention toward methods that generate videos by animating targets in the source image to mimic the motion of corresponding targets in the driving video, as illustrated in Fig.~\ref{fig: method intro}. \cite{FOMM} detects the keypoint position of targets in a source image and driving video, calculates the affine transformation between corresponding keypoints, and synthesizes the new images by means of a GAN. \cite{siarohin2021motion} extracted meaningful and consistent regions and models non-object-related global motion, with an additional affine transformation to force decoupling between the foreground and background (BG). \cite{TPS} uses thin-plate spline motion estimation to produce a more flexible optical flow to warp the feature maps of the source image to the feature of the driving video and leverage multi-resolution occlusion masks. \cite{DAM} uses motion anchors to capture both appearance and motion information and a latent root anchor to link the motion anchors to enhance representation learning of the structural information of objects. \par
Although the aforementioned methods have achieved significant success in the synthesis of natural videos, their applicability to US video synthesis may be limited. This discrepancy arises from fundamental differences in imaging principles between the natural and medical contexts. For example, Gaussian noise is widespread in natural images, whereas Poisson noise is more common in medical images. Natural images typically consist of color representations, whereas US images are primarily grayscale, with different tissue types and structures delineated using varying shades of gray. Consequently, the synthesis of US videos with sufficiently high fidelity and accurate animation to the highly dynamic motion in anatomic structures remains a significant challenge.\par
To synthesize high-quality US videos, careful attention must be paid to the following key considerations:
\textbf{First}, the integrity of anatomic structures must be preserved, and their motion must be simulated accurately. The anatomic structures in US images typically display complex and uncertain boundaries, making it challenging to maintain their integrity and mimic their highly dynamic motion.
\textbf{Second}, a balance must be achieved between preserving speckle noise and maintaining the textural information. Speckle noise frequently manifests in US images, whereas textural information plays a vital role in medical image analysis \cite{wei2020benigntexture}. However, US images often exhibit a grainy or speckled appearance owing to speckle noise, which makes it difficult to accurately maintain the textural information \cite{choi2020despeckling}.
\textbf{Third}, a common limitation of most existing video synthesis methods \cite{FOMM,TPS,DAM} is that they are typically based on the synthesis of individual frames, which might not ensure the continuity and consistency of content between frames in the synthesized video. \par
In this paper, we present a novel online feature-decoupling framework called \ourmodel~for high-fidelity US video synthesis. This is the first study that synthesizes US videos by animating a static source image using the motion provided by the driving video (Fig.~\ref{fig: method intro}). Our primary contributions are fourfold:
\begin{itemize}
    \item We integrate anatomic information into keypoint learning using a weakly supervised approach, aiming to enhance the preservation of anatomic integrity and achieve more accurate replication of anatomic motion in driving videos, all while minimizing labeling burdens.
    \item By disentangling content and texture features, the proposed dual-decoder generator enables more precise capture of US image features, resulting in better preservation of both content and textural information.
    \item A multi-scale discriminator is utilized to extract comprehensive local and global visual cues from images of varying scales, promoting consistency between the distributions of real and generated videos, and enhancing the sharpness and fine details of synthesized videos.
    \item We propose an innovative online learning strategy to constrain the motion trajectory of keypoints in the generated video, which enhances the temporal connections between video frames, making the generated video more fluid.
\end{itemize}\par
\begin{figure*}[!t]
    \centering
    \includegraphics[width=0.9\linewidth]{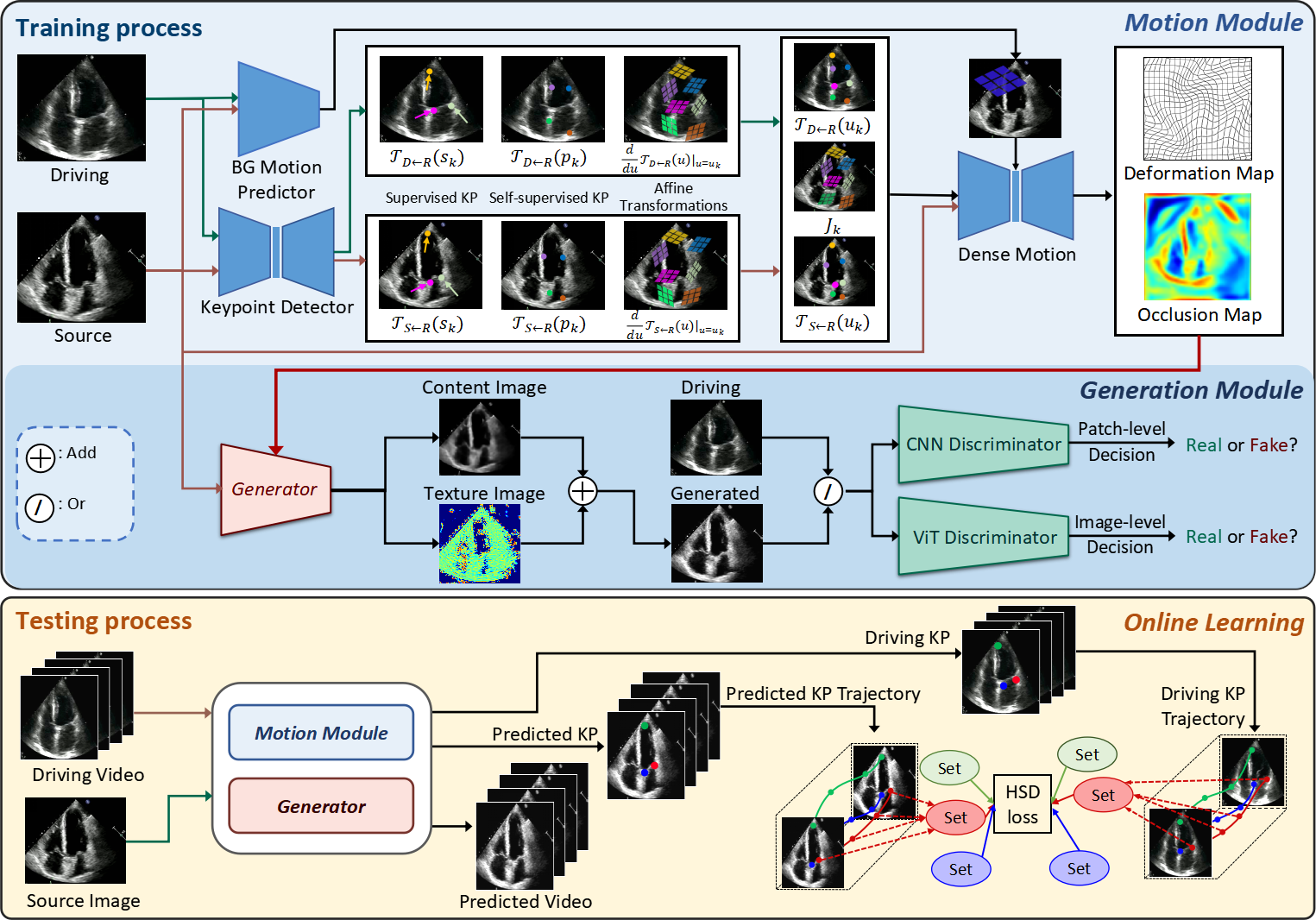}
    \caption{Overview of online feature decoupling framework (\ourmodel). Texture image and occlusion map are gray maps visualized in pseudo-color.}
    \label{fig:framework}
\end{figure*}
The proposed methodology extends upon our MICCAI work \cite{OUR} with the following efforts:
First, to simulate the affine transformation more accurately, an additional BG affine transformation is used to estimate the non-object-related global motion. Second, a vision transformer (ViT) \cite{VIT} is incorporated into the discriminator to additionally extract features from images of varying scales, encouraging the generator to focus on detailed information. Third, the motion trajectories of anatomic structures in videos synthesized during online learning are constrained to enhance the coherence between video frames. We also included two in-house echocardiograph cardiography datasets for comprehensive validation.
\begin{figure}[!t]
    \centering \includegraphics[width=1\linewidth]{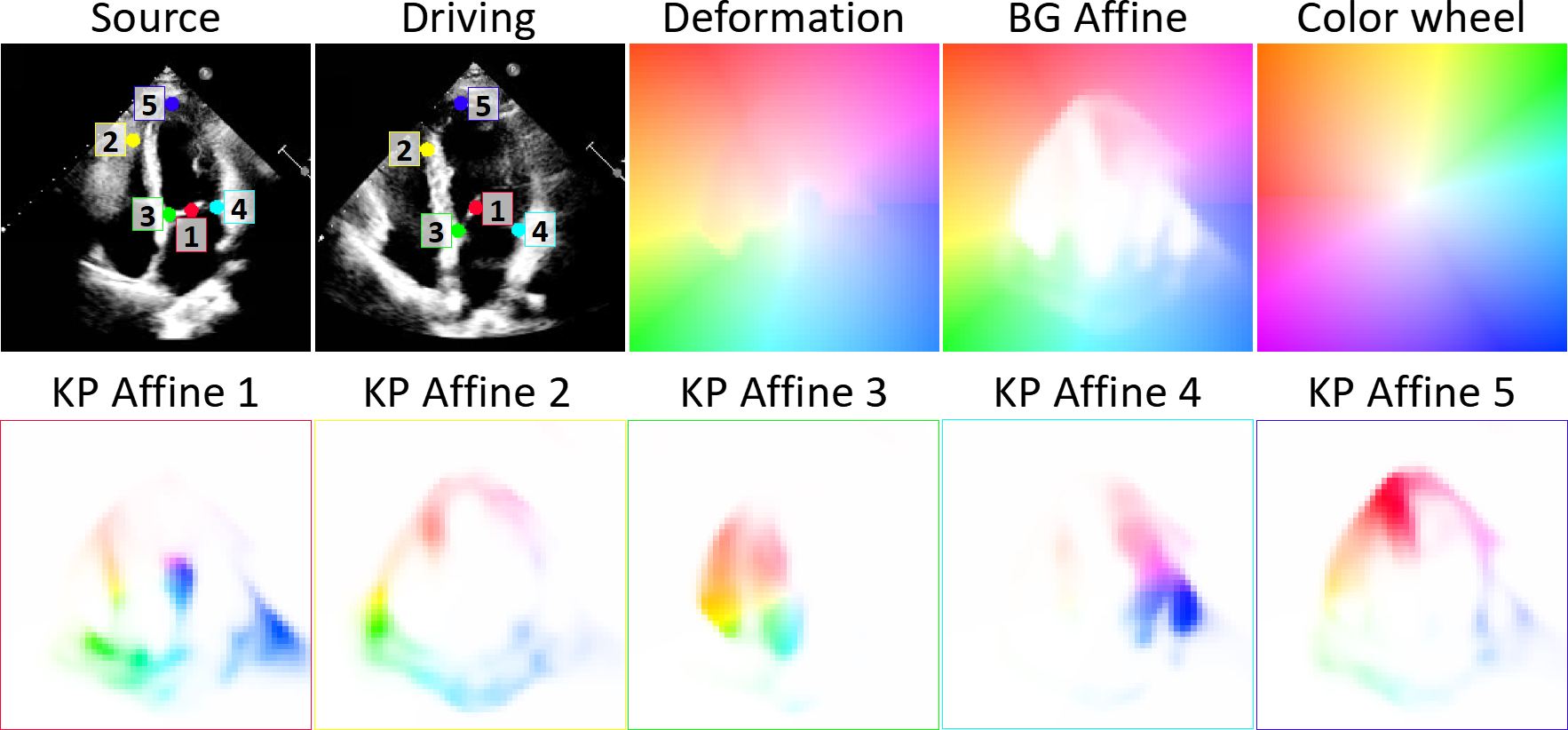}
    \caption{Visualization of a deformation map, which is the sum of all individual affine maps.}
    \label{Fig:intermidia_visual}
\end{figure}
\section{Methodology}
The overall framework of \ourmodel~is illustrated in Fig.~\ref{fig:framework}. During the training process, the keypoint detector predicts the location and parameters of the affine transformation (KP Affines in Fig.~\ref{Fig:intermidia_visual}) on K pairs of keypoints. Simultaneously, the BG motion predictor estimates the non-object-related global motion using an additional affine transformation (BG Affine, Fig.~\ref{Fig:intermidia_visual}). Subsequently, the dense motion network calculates the deformation and occlusion maps, and the generator then synthesizes the final US frame. Finally, the generated and driving frames are sent to the discriminator for identification. During the testing process, we first generated a US video and predict the keypoints of the driving and generated videos. We then calculated the Hausdorff distance (HSD) \cite{HSD} between the keypoint sets of the two videos. Finally, we used the HSD loss to optimize the keypoint detector, generator, and BG motion predictor. Following sections will introduce \ourmodel~ in details.
\subsection{Weakly-Supervised Training for Motion Estimation}
As in FOMM \cite{FOMM}, we use dense deformation fields $\mathcal{T}_{S \leftarrow D}$ from frame $D$ in the driving video to source image $S$ to transfer motion from $D$ to $S$. A Taylor expansion is used to represent $\mathcal{T}_{S \leftarrow D}$ by a set of keypoint locations and the first-order motion information ($\frac{d}{du}$ in Fig.~\ref{fig:framework}) between corresponding keypoints. Thus, the localization of keypoints is crucial for motion transformation. The first-order motion information between corresponding keypoints (KP Affine) is illustrated in Fig.~\ref{Fig:intermidia_visual}. Given the transformation and assumed reference frame $R$, a Jacobian matrix ($J_k$ in Fig.~\ref{fig:framework}) can be obtained to predict the motion between $S$ and $D$. \par
In this study, keypoints were detected by the keypoint detector. Although a fully supervised training strategy can enhance the accuracy of keypoint localization, it is difficult to obtain medical US video datasets with densely annotated anatomic structure keypoints, as manual annotation of such datasets can be time-consuming and expensive. To address this limitation, we employed a weakly-supervised training strategy to train the keypoint detector. Specifically, building on the self-supervised method \cite{FOMM}, we introduced a small number of supervised keypoints to incorporate essential anatomic information into the training process. This approach leverages the benefits of both self-supervised learning and limited supervised annotation, enabling effective training of the keypoint detector while eliminating the need for extensive manual labeling. \par
\begin{figure*}[!t]
\centering
\includegraphics[width=.9\linewidth]{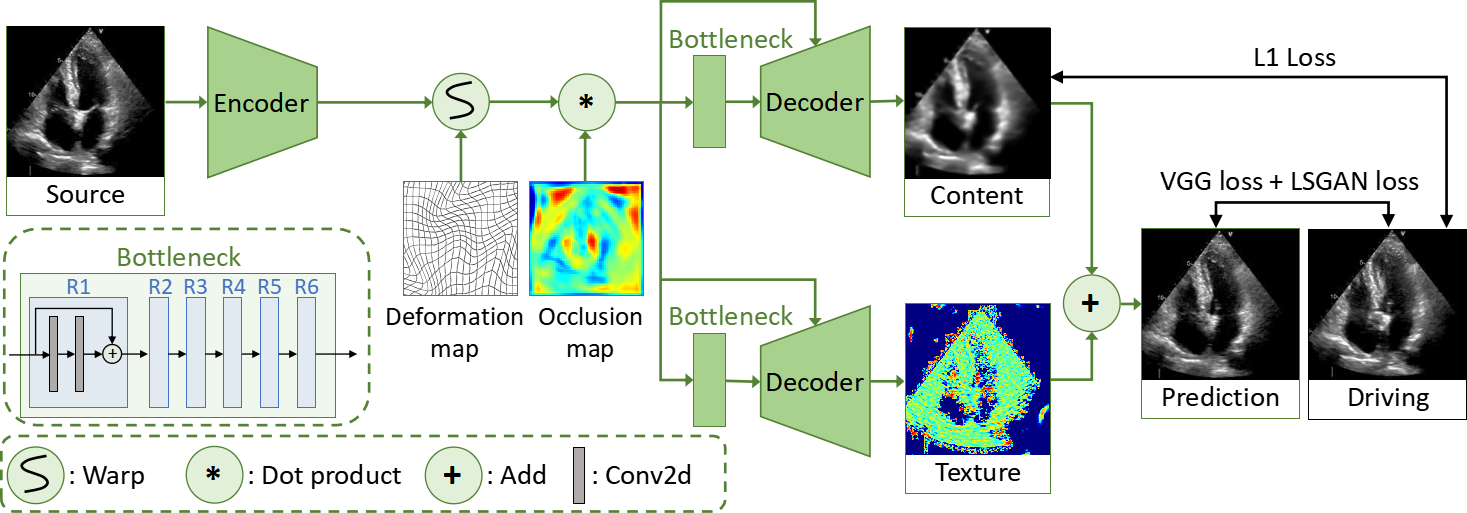}
\caption{Structure of dual-decoder generator. R1-R6 are the same residual blocks. Texture image and occlusion map are gray maps visualized in pseudo-color.}
\label{fig:generator_detail}
\end{figure*}
Among all keypoints, the self-supervised component was learned through thin-plate spline deformations (refer to \cite{FOMM}), and the fully supervised part was learned via manual annotation.
For self-supervised keypoint learning, the equivariance losses $L_{eq}$ in terms of displacements $L_{eq1}$ and affine transformations $L_{eq2}$ are calculated as follows:\par
\begin{equation}
\label{equa:eq1} 
L_{eq1}=\left \| \mathcal{T}_{X\leftarrow R}(p_k) - \mathcal{T}_{X\leftarrow Y} \circ \mathcal{T}_{Y\leftarrow R}(p_k)\right \|_1,
\end{equation}
\begin{equation}
\label{equa:eq2}
L_{eq2}=\left \| 1 - {({\mathcal{T}_{X\leftarrow R}^{'}(p)|}_{p=p_k}){(\mathcal{T}_{X\leftarrow Y}^{'})}{(\mathcal{T}_{Y\leftarrow R}^{'}(p)|_{p=p_k})}}\right \|_1,
\end{equation}
\begin{equation}
\label{equa:eq2_a}
\mathcal{T}_{X\leftarrow Y}^{'} = {(\mathcal{T}_{X\leftarrow Y}^{'}(p)|_{p= \mathcal{T}_{Y\leftarrow R}(p_k)})},
\end{equation}
\begin{equation}
\label{equa:eq_total}
L_{eq} = L_{eq1} + L_{eq2},
\end{equation}
where $X$ and $Y$ denote the driving/source and the frames following the thin-plate spline transformation of $X$, respectively, and $p_k$ denotes the $k_{th}$ self-supervised points.
For the supervised component, L2 loss is used to constrain the differences between true and predicted heatmaps, and can be calculated as
\begin{equation}
\label{equa:lkey}
L_{key} = \left \| \mathcal{T}_{R\leftarrow X}(s_k) - Heatmap(s_k)\right \|_2,
\end{equation}
where $s_k$ denotes the $k_{th}$ supervised keypoint. We then use the dense motion network proposed by \cite{FOMM} to learn the dense deformation fields and occlusion maps. Deformation fields provide deformation information, whereas occlusion maps provide vital indications to inform the model where to focus.\par
\subsection{Dual-Decoder for Content and Texture Decoupling} 
Content and texture are two vital elements in US images, providing important information about the structure and composition of imaged tissue. The content of US frames represents the underlying anatomic structures being imaged, whereas texture reflects the visual appearance of tissues and represents additional diagnostic information that can help identify abnormalities \cite{rezazadeh2022explainable}. However, the synthesis of textural features in US videos is complexified by speckle noise. \par
Previous studies \cite{FOMM,TPS,DAM} primarily focused on synthesizing natural videos, wherein both content and texture are learned under a unified path. However, a unified path architecture may lead to the potential loss of high-frequency textural information, resulting in video blurring. By decoupling the content and texture features using a dual decoder in the generator, we can generate US videos that more accurately retain the content and textural information of US frames. \par
Fig.~\ref{fig:generator_detail} illustrates the structure of a dual-decoder generator that accepts the source, deformation, and occlusion maps as input. The intermediate features extracted by the encoder from the source image learn deformation information provided by deformation and occlusion maps through warp and dot products. The deformed intermediate features are then sent to the bottleneck and each upsampling layer of the decoder to strengthen the learning process. The final predicted image is obtained by summing the pixel values of the content and texture images. This learning mode is driven by carefully designed loss functions. \par
For content learning, we utilized the L1 reconstruction loss, which enforces pixel-level consistency between the driving frame and predicted content. This loss enables the generator to accurately capture content information from the driving frame. Because ground truth (GT) texture is not available for texture learning, we leveraged the feature reconstruction VGG loss \cite{johnson2016perceptual}. VGG loss operates in a high-level feature space and encourages similarity between the driving and generated frames in terms of textural representations. By combining these two loss functions, we ensured pixel-level consistency and texture fidelity in the synthesized US video frames, resulting in high-quality, visually coherent output. The two losses were calculated at multiple resolutions obtained by downsampling operations including $256\times256$, $128\times128$, $64\times64$ and $32\times32$ which can be expressed as follows:
\begin{equation}
\label{equa:RECL1}
L_{recL1} = \sum_{i=0}^I \left \| Down_i(D) - Down_i(G_c(S))\right \|_1,
\end{equation}
\begin{equation}
\label{equa:RECVGG}
L_{recVGG} = \sum_{i=0}^I \sum_{j=0}^J \left \| V_j(Down_i(D)) -V_j(Down_i(G_f(S)))\right \|_1,
\end{equation}
where $Down_i$ is the $i_{th}$ downsample, $V_j$ is the $j_{th}$ activation layer of VGG network, $G_c$, $G_f$, $D$, and $S$ are the content image, final prediction, driving, and source, respectively.
\begin{figure*}[!t]
\centering
\includegraphics[width=.9\linewidth]{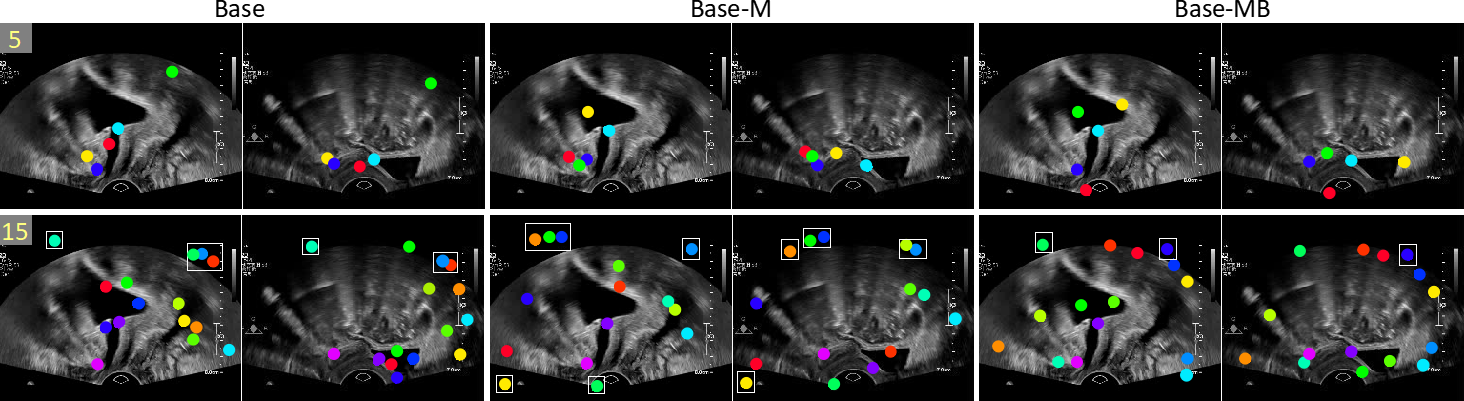}
\caption{Keypoints estimated by Base, Base-M, and Base-MB. Base represents our previously developed model \cite{OUR}. Base-M and Base-MB represent its extensions that gradually incorporate the multi-scale discriminator ('-M') and BG motion predictor ('-B'). 5 and 15 refer to keypoint numbers. White boxes indicate keypoints in the black BG. }
\label{fig:model_kp_visual}
\end{figure*}
\subsection{Adversarial Learning with a Multi-Scale Discriminator.}
Although the aforementioned designs have successfully addressed the challenges of estimating motion, content, and texture in synthesized videos, a limitation remains in generating finer details. To mitigate this limitation, we introduce an adversarial training strategy \cite{goodfellow2020generative} to further enhance the sharpness and fine details of generated videos by adding a multi-scale discriminator to our framework. \par
Previous results in the field of neural architecture \cite{raghu2021vision} indicate that supervised and self-supervised representations are similar, whereas the ViT and convolutional neural network (CNN) are different. Compared with the conventional CNN, the ViT exhibits a higher degree of similarity between representations obtained at the shallow and deep layers and retains more spatial information. Accordingly, the discriminator leverages ViT and CNN to extract comprehensive local and global representations from images of varying scales. This enhances the realism of generated US videos by minimizing the distribution distance between generated and real videos across a broader feature space.\par
We fed real and synthesized images of varying scales to the discriminators. The multi-scale images can be obtained by downsampling operations. The CNN discriminator attempts to classify whether each $N \times N$ patch in an image is real or fake, penalizes the structure only at the patch scale, and obtains its output by averaging the responses for all patches. The ViT discriminator obtains classification results based on the entire image. To produce stable and high-quality samples and alleviate mode collapse, we used the LSGAN loss \cite{LSGAN} as the adversarial loss to train the generator ($L_{G}^{LSGAN}$) and discriminators ($L_{Dis}^{LSGAN}$).\par
\begin{equation}
\label{equa:LG}
L_{G}^{LSGAN} = E[(Dis(G_f(S)-1)^2],
\end{equation}
\begin{equation}
\label{equa:Dis}
L_{Dis}^{LSGAN} = E[(Dis(D)-1)^2] + E[(Dis(G_f(S))^2].
\end{equation}
Here, $G_f$ denotes the final prediction, and $S$ and $D$ indicate the source and driving frame, respectively.We adopted feature matching loss \cite{wang2018high} to encourage similar
intermediate representations between generated and driving in discriminators. The feature matching loss can be expressed by Eq.~\ref{equa:feat}:
\begin{equation}
\label{equa:feat}
L_{feat} = \sum_{n=0}^N \left \| Dis_n(D) -Dis_n(G_{f}(S)))\right \|_1,
\end{equation}
$Dis_n$ denotes the $n_{th}$-intermediate output of the discriminator.
\subsection{Background Motion Estimation}
After applying the multi-feature discriminator, we observed that the predicted locations of keypoints were more closely distributed around anatomic structures when the number of keypoints was five, and more dispersed throughout the whole image when the number of keypoints was 15. Specifically, same as the Base case, some keypoints predicted by Base-M appeared in the black BG region as the number of keypoints number increased (Fig.~\ref{fig:model_kp_visual}). This is because the dense motion network overlooks global motion when computing the deformation and occlusion maps. Consequently, when the number of keypoints increases, the keypoint predictor compensates for estimating BG motion by predicting the points in the black BG. To minimize the occurrence of such keypoints, we introduced an additional BG affine transformation similar to \cite{siarohin2021motion} to estimate non-object-related global motion. 
As shown in Fig.~\ref{Fig:intermidia_visual}, the BG affine primarily focuses on estimating the black BG area transformation, distinguishing it significantly from KP affines. The BG affine, along with the KP affine were combined to form the final dense deformation fields. The BG affine transformation was calculated using the BG motion predictor, which receives the combination of the source and driving frames as input, and outputs the BG affine transformation $\mathcal{T}_{D\leftarrow S}$ as follows:
\begin{equation}
 \mathcal{T}_{D\leftarrow S} =
\begin{bmatrix} 
t_1 & t_2 & t_3\\
t_4 & t_5 & t_6
\end{bmatrix},
\end{equation}
where $t_1, t_2,...,t_6$ are constants. Fig.~\ref{fig:model_kp_visual} illustrates the effects of incorporating the BG motion predictor. Notably, with the inclusion of this predictor, the predicted keypoints are more distributed outside the black BG area.\par
\begin{table*}[!htbp]
\centering
\caption{Comparison with state-of-the-art and ablation study results on three different datasets. $*$ denotes passed \textit{t-test, $p < 0.05$}}
    \begin{tabular}{c|l|lccccc|lc}
        \toprule 
          \multirow{2}{*}{Dataset}& \multirow{2}{*}{Method}& \multicolumn{6}{c}{Reconstruction}  & \multicolumn{2}{c}{Animation}\\
           \cline{3-10}   
           & &FVD$\downarrow$&FID$\downarrow$&LPIPS$\downarrow$ &L1$\downarrow$ &PSNR$\uparrow$&SSIM$\uparrow$ &FVD$\downarrow$ &FID$\downarrow$ \\
           \hline
           \multirow{7}{*}{A4C}& FOMM \cite{FOMM}& $285.85^*$ & 51.43& 0.0461 & {\color{blue}0.0184} & 27.31 & 0.8370 & $450.14^*$ & 92.96\\ 
        & TPS \cite{TPS} & $313.75^*$ & 55.45 & 0.0421&0.0190 & 27.48 & 0.8388
& $545.84^*$ & 105.87  \\ 
        & DAM \cite{DAM}& $1181.66^*$ & 238.59 &0.2690& 0.0636 & 16.79 & 0.6173 & $1216.94^*$ & 242.83\\ 
        & Base & $272.55^*$ & 29.89& 0.0338 & 0.0189 & 27.29 & 0.8230 & $448.94^*$ & 78.29\\ 
         & Base-M & $243.62^*$ & {\color{blue}29.59} &$0.0329$ & 0.0189 & 27.14 & $0.8352$ & $415.31^*$ & {\color{blue}72.43}\\ 
       &   Base-MB & 228.75 & 30.83 &0.0319 & 0.0187 & 27.33& 0.8373 & 407.19 & 76.74\\ 
        &  \ourmodel & {\color{blue} 221.28}  & 29.96 &{\color{blue}0.0319} & 0.0186 & {\color{blue}$27.51$} & {\color{blue}0.8394} & {\color{blue}401.45} & 76.01\\ 
            \hline
            \multirow{7}{*}{A2C}& FOMM \cite{FOMM}& $324.96^*$ & 63.01 &0.0650 & 0.0195 & 26.94 & 0.8189 & $499.09^*$ & 149.42 \\ 
                &TPS \cite{TPS}& $329.46^*$ & 59.04 & 0.0618 & 0.0199 & 26.97 & 0.8245 & $639.83^*$ & 164.30\\ 
                &DAM \cite{DAM} & $641.60^*$ & 74.40 & 0.0892 & 0.0260 & 22.50 & 0.6335  & $839.06^*$ & 245.37 \\ 
                & Base & $289.27^*$ &  {\color{blue} 35.65} &0.0484 &  0.0198 & 26.98 &  0.8232 & $495.98^*$ & 135.57\\ 
                &Base-M & $276.89^*$ & 39.00 & 0.0486 & 0.0193 & $27.08$ & $0.8235$ & $474.55^*$ & 133.29\\ 
                &Base-MB & 243.07 & 36.15& $0.0483$ & 0.0194 & 27.05 & 0.8251 & $463.47^*$ &128.72\\ 
                &\ourmodel & \color{blue}237.13 & 35.98 &{\color{blue}0.0482} & {\color{blue}0.0192} &   {\color{blue}27.11} & {\color{blue} $0.8278$} &  {\color{blue}452.97} & {\color{blue}128.61}\\ 
           \hline
            \multirow{7}{*}{Pelvic}& FOMM \cite{FOMM} & $197.61^*$ & 71.13 &0.058 & 0.0232 & 26.54 & 0.7679 & $621.57^*$ & 92.56\\ 
            &TPS \cite{TPS} & $178.41^*$ & 61.53 & 0.052 & 0.0227 & 26.78 & 0.7964 &$476.25^*$& 89.82 \\ 
            &DAM \cite{DAM}& $300.32^*$ & 61.13 & 0.097 & 0.0224 & 26.93 & 0.7791 & $469.73^*$ & 79.18 \\ 
            &Base & $195.96^*$ & 61.33 & 0.060 & 0.0212 & 27.11 & 0.7991 & $576.16^*$ & 72.26\\ 
            &Base-M & $174.28^*$ & 61.21 & $0.057$ & 0.0211 & $26.80$ & 0.7990 & $401.94^*$ & 58.12\\ 
            &Base-MB & $164.05^*$ & {\color{blue}60.02} & 0.057 & 0.0211 & 26.96 & 0.8028 & $357.99^*$ & {\color{blue}52.32}\\ 
           & \ourmodel & {\color{blue}152.15} & 60.26 & {\color{blue}0.042} & {\color{blue}0.0210} &  {\color{blue}27.13} & {\color{blue}0.8076} & {\color{blue}334.45} & 52.78 \\
           \bottomrule 
     \end{tabular}
 \label{tab:COMPARISION}
\end{table*}
\subsection{Online Keypoint Motion Trajectory Constraint}
Our approach synthesizes US videos on a frame-by-frame basis, often leading to temporal discontinuities in the motion of anatomic structures. Specifically, the deformation map of each frame in the generated video is calculated from keypoints between the corresponding driving frame and source image, but the prediction of keypoints is unstable. To solve this problem, we propose an online learning process where the motion trajectories of keypoints in generated videos are constrained according to the driving videos using the HSD \cite{HSD}.\par
The HSD is widely used as a measure of dissimilarity between two trajectories, wherein each trajectory is represented by a set of points. For two nonempty point sets $\mathbb{A}$ and $\mathbb{B}$, the maximum value of the minimum distance of a point in $\mathbb{A}$ with respect to all points in $\mathbb{B}$ is a one-way HSD from $\mathbb{A}$ to $\mathbb{B}$:
\begin{equation}
h(\mathbb{A}, \mathbb{B})=\mathop{max}\limits_{a\in A}\mathop{min}\limits_{b\in B}d(a,b),
\end{equation}
where $d(a, b)$ denotes the Euclidean distance between points $a$ and $b$. Similarly, the one-way HSD from $\mathbb{B}$ to $\mathbb{A}$ is
\begin{equation}
h(\mathbb{B}, \mathbb{A})=\mathop{max}\limits_{b\in B}\mathop{min}\limits_{a\in A}d(a,b).
\end{equation}
A one-way HSD is typically used to represent the maximum deviation of a set relative to another set. The larger of two one-way HSDs is called the two-way HSD between $\mathbb{A}$ and $\mathbb{B}$ and is usually denoted as
\begin{equation}
H(\mathbb{A}, \mathbb{B})=\mathop{max}\left\{h(\mathbb{A}, \mathbb{B}),h(\mathbb{B}, \mathbb{A})\right\}.
\end{equation}
In this study, a two-way HSD was used to indicate the distance between the motion trajectories of corresponding keypoints. \par
In the testing phase (see the testing process in Fig~\ref{fig:framework}), we first generated a US video based on the driving video and source image. Subsequently, we obtained keypoint sets $\mathbb{G}_n=\left\{g_n^1, g_n^2,...,g_n^f\right\}$ for the predicted video and $\mathbb{P}_n = \left\{p_n^1, p_n^2,...,p_n^f\right\}$ for the driving video, where $n$ is the $n_{th}$ keypoint and $f$ is the frame number. Then, we calculated $L_{HSD}$ using Eq.~\ref{eq:lhs}. Finally, we used $L_{HSD}$ to optimize the keypoint detector, generator, and BG motion predictor. We repeated the previous step 10 iterations and selected the generated US video with the lowest FVD as the final result. \par
\begin{equation}
L_{HSD} =\sum_{n=0}^N HSD {(\mathbb{P}_n, \mathbb{G}_n)},
\label{eq:lhs}
\end{equation}
where $N$ is the total number of keypoints.
\section{Experiments and Results}
\subsection{Experimental Setup}
\begin{figure}[!t]
\centering
\includegraphics[width=0.92\linewidth]{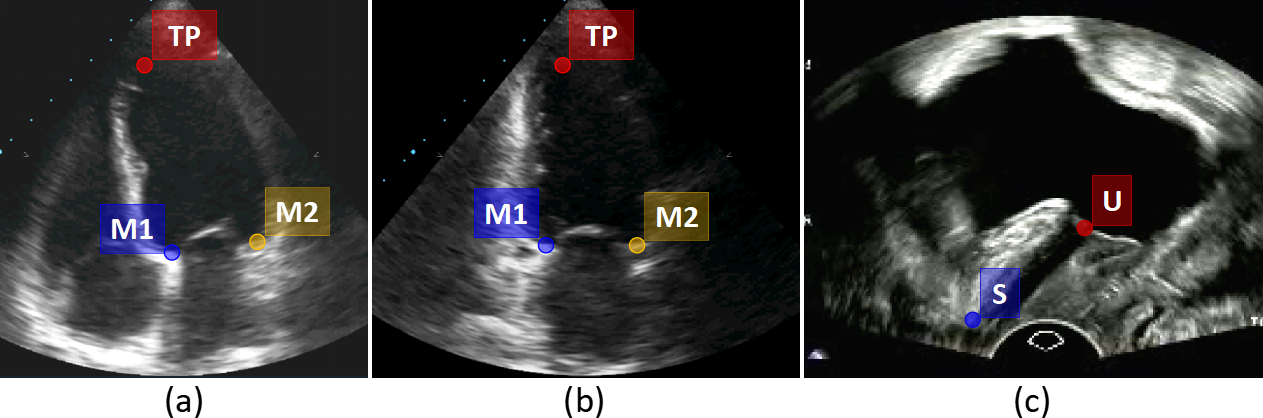}
\caption{Annotated keypoints of our datasets.}
\label{fig:point_visual}
\end{figure}
\textbf{Datasets.} 
This study was approved by local institutional review boards. We trained and validated methods on A4C heart echocardiography, A2C heart echocardiography, and pelvic floor US video datasets. The \textbf{A4C} heart echocardiography dataset contains 159 training and 65 testing videos. As shown in Fig.~\ref{fig:point_visual} (a), three keypoints were annotated. Each video spans 10--209 frames. The \textbf{A2C} heart echocardiography dataset contains 137 training and 57 testing videos. Three keypoints were annotated (Fig.~\ref{fig:point_visual} (b)) in each frame. Each video spans 10--114 frames. The \textbf{Pelvic Floor} dataset contains 134 training and 35 testing videos. Two keypoints were annotated as shown in Fig.~\ref{fig:point_visual} (c). Each video spans 37--88 frames. All the frames were resized and zero-padded to $256\times256$. All the keypoints were annotated by experts using the Pair software \cite{Pair}.\par
\textbf{Evaluation Metrics.}
We employed six metrics to evaluate the proposed method in terms of detail preservation, image and video fidelity, feature similarity, and structural similarity.\par
$\bullet$ L1 loss measures the level of detail captured by the generated image compared to the source image at pixel level.\par
$\bullet$ Frechet Inception Distance (FID) \cite{FID} assesses the distance of the feature vector $f \in \mathbb{R}^{N \times 2048}$ extracted by Inception \cite{szegedy2017inception} between $N$ generated image $g$ and real image $r$:
\begin{equation}
\label{eq:fid}
FID(g,r)={\left \| \mu_g-\mu_r\right \|}_2^2 + T_r \left(\Sigma_g + \Sigma_r - {2\left( \Sigma_g \Sigma_r \right )} ^\frac{1}{2} \right),
\end{equation}
where $\mu_g$/$\mu_r$ and $\Sigma_g$/$\Sigma_r$ are the mean values and covariance matrices of each vector, respectively, and $T_r$ denotes the sum of the elements of the main diagonal of the matrix.\par
$\bullet$ Frechet Video Distance (FVD) \cite{FVD} evaluates temporal coherence in generated videos and assesses how well the motion aligns with that in driving videos. The computation of FVD is similar to that of FID, except the I3D \cite{I3D} is used to extract video features.\par
$\bullet$ Learned Perceptual Image Patch Similarity (LPIPS) \cite{LPIPS} analysis the similarity between the generated image $x$ and the driving frame $x_0$ as follows:\par
\begin{equation}
\label{eq:lpips}
LPIPS(x,x_0)=\sum_{l} \frac{1}{H_lW_l} \sum_{h,w} {\left \| w_l \textstyle\bigodot \left ({\hat{y}}_{hw}^l - \hat{y}_{0hw}^l \right) \right \|}_2^2.
\end{equation}
The feature stack $y^l, \hat{y}^l \in {\mathbb{R}}^{H_l\times W_l \times C_l}$ is extracted from the $l$ layer of feature extract model. Vector $w_l = 1 \forall l$, $w_l \in {\mathbb{R}}^{C_l}$.\par
$\bullet$ Structural Similarity Index Measure (SSIM) \cite{SSIM} measures structural similarity between generated image $x$ and driving frame $y$, and can be calculated as follows:\par
\begin{equation}
\label{eq:ssim}
SSIM(x,y)={\left[l\left(x,y\right) \right]}^\alpha {\left[c\left(x,y\right) \right]}^\beta {\left[s\left(x,y\right) \right]}^\gamma,
\end{equation}
\begin{equation}
\label{eq:ssim_l}
{l\left(x,y\right)}=\frac{2\mu_x\mu_y + c_1}{{\mu_x^2}+{\mu_y^2}+{c_1}},
\end{equation}
\begin{equation}
\label{eq:ssim_c}
{c\left(x,y\right)}=\frac{2\sigma_{xy} + c_2}{{\sigma_x^2}+{\sigma_y^2}+{c_2}},
\end{equation}
\begin{equation}
\label{eq:ssim_s}
s\left(x,y\right) = \frac{\sigma_{xy}+c_3}{{\sigma_x\sigma_y}+{c_3}},
\end{equation}
where $\alpha > 0$, $\beta > 0$, $\gamma > 0$. 
${l\left(x,y\right)}$, ${c\left(x,y\right)}$, and $s\left(x,y\right)$ are the comparison of luminance, contrast, and structure, respectively. $\mu_x$/$\mu_y$ and $\sigma_x$/$\sigma_y$ represent the average values and standard deviation, respectively. $\sigma_{xy}$ represents the covariance between $x$ and $y$. $c_1$, $c_2$, and $c_3$ are constant.\par
$\bullet$ Peak Signal-to-Noise Ratio (PSNR) quantifies the noise and distortion level of the generated image.\par
\textbf{Implementation Details.}
We implemented our method in Pytorch and trained the system with the Adam optimizer for 100 epochs using multiple GPUs including NVIDIA GTX 2080Ti with 12GB for the pelvic floor dataset and NVIDIA GTX 3090 with 24GB for heart echocardiography datasets. The batch size was set to eight for NVIDIA GTX 3090 and four for NVIDIA GTX 2080Ti, and the learning rate was set to 0.0002. The number of keypoints was set to 15 for the A4C heart echocardiography dataset and 10 for the other datasets. The keypoint detection and dense motion network employed a U-Net \cite{UNet} structure with five downsampling and upsampling blocks. The bottleneck contained six residual blocks with two convolutional layers. The BG motion predictor, ResNet18 \cite{RESNET} with a changed final linear layer, outputs six constants. For the CNN discriminator, we implement the PatchGAN \cite{isola2017image} structure with four convolutional layers. For the ViT discriminator, we employed a base ViT with a patch size of $16 \times 16 $. The weights of the losses $L_{eq}$, $L_{key}$, $L_{recL1}$, $L_{recVGG}$, $L_{G}^{LSGAN}$, $L_{Dis}^{LSGAN}$, $L_{feat}$, and $L_{HSD}$ were 10, 100, 10, 10, 1, 1, 10, and 1, respectively.\par
\subsection{Comparison With Existing Methods}
\begin{figure*}[!t]
    \centering
    \includegraphics[width=0.92\linewidth]{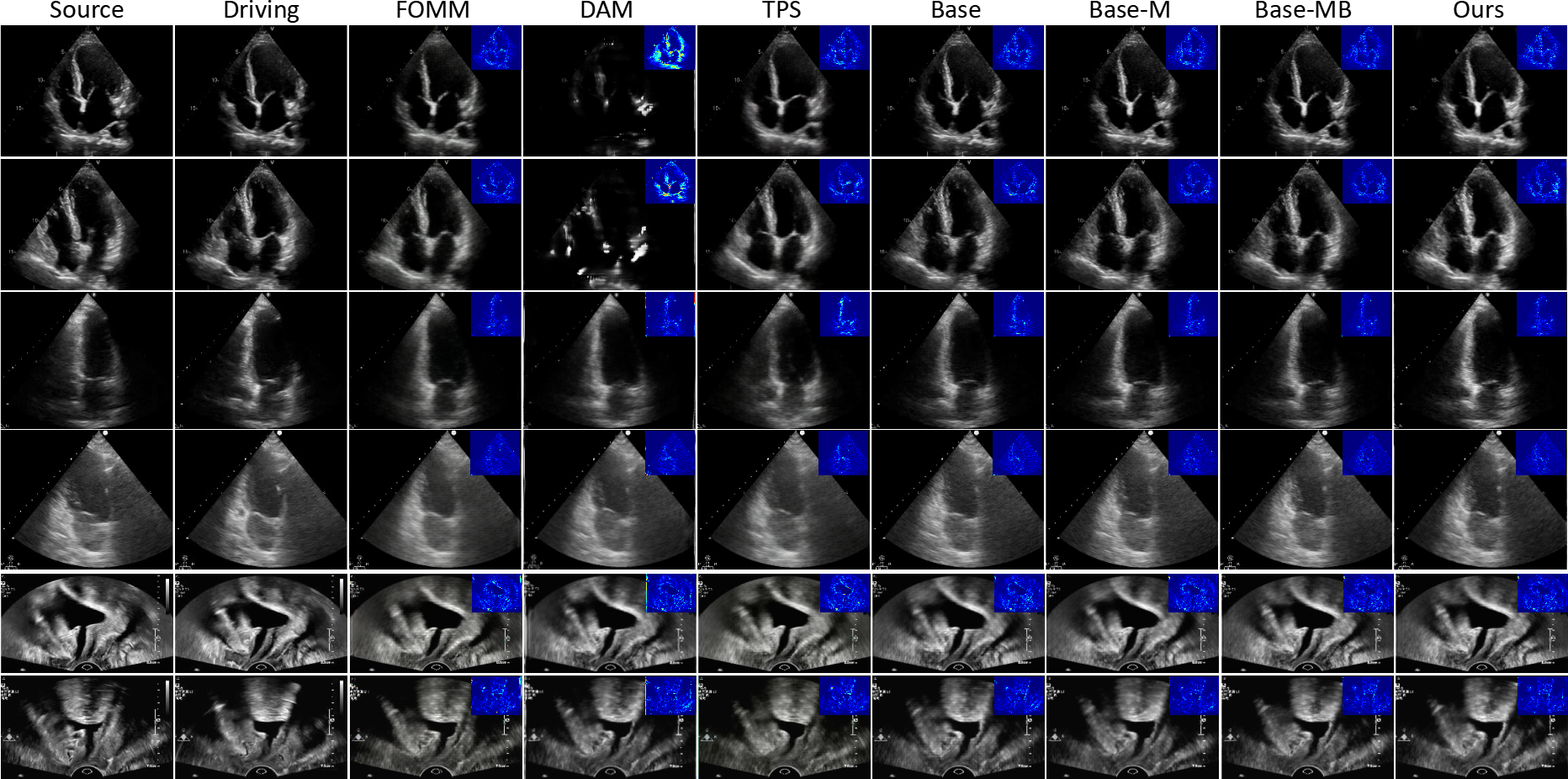}
    \caption{Qualitative comparison on reconstruction task. The top two rows represent A4C heart echocardiography, the middle two rows represent A2C heart echocardiography, and the bottom two rows represent the pelvic floor. The top right corner of each synthesized image shows a difference heatmap.}
    \label{fig:reconstraction_result}
\end{figure*}
\begin{figure*}[!t]
    \centering
    \includegraphics[width=0.92\linewidth]{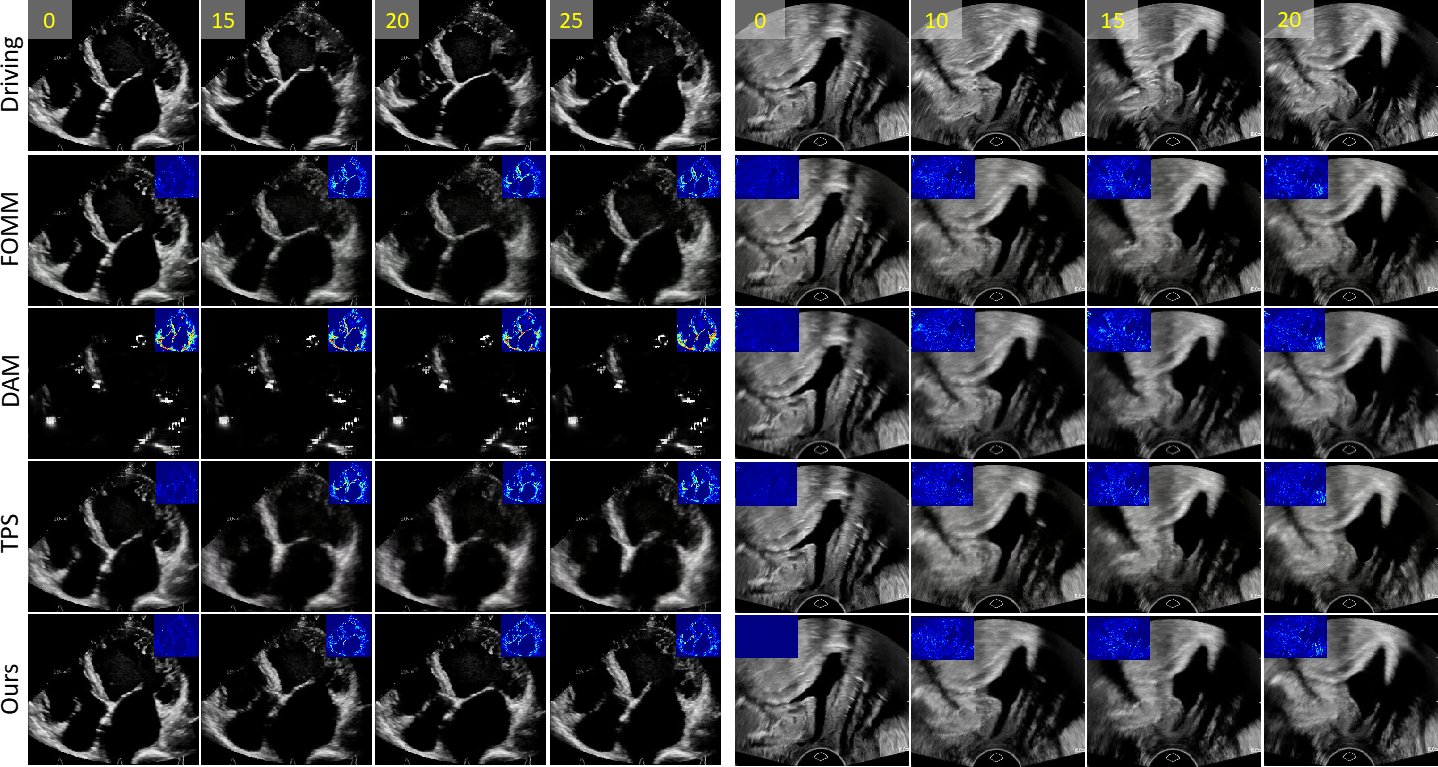}
    \caption{Visualization of serial frames of the same video reconstructed by different methods. The frame number alternated between 0, 15, 20, and 25. The top right corner of each synthesized A4C heart echocardiography frame and the top left corner of each synthesized pelvic floor frame display different heatmaps.}
    \label{fig:reconstraction_fluid}
\end{figure*}
We evaluated the performance of \ourmodel~on two tasks: reconstruction and animation. For the reconstruction task, we designated videos from the testing set as driving and their first frames as source. For future clinical use, we can collect US videos of relevant cases for training to better synthesize videos of corresponding diseases. Meanwhile, we can use random US images as source and videos of such cases as driving to synthesize such videos. Therefore, for the animation task, we randomly selected a video from the training set as the driving and an intermediate frame in another video from the testing set as the source. Because the GT was the driving video in the reconstruction task and was unavailable in the animation task, we utilized all six metrics to evaluate the former and only considered FID and FVD for the latter.\par
We compared our method with the state-of-the-art motion transfer methods, TPS \cite{TPS} and DAM \cite{DAM}, and the classic motion transfer method FOMM \cite{FOMM}. To ensure a fair comparison, we retrained all methods with the same training schedule and standardized the source and driving conditions during the test.\par
\begin{figure*}[!t]
    \centering
    \includegraphics[width=0.92\linewidth]{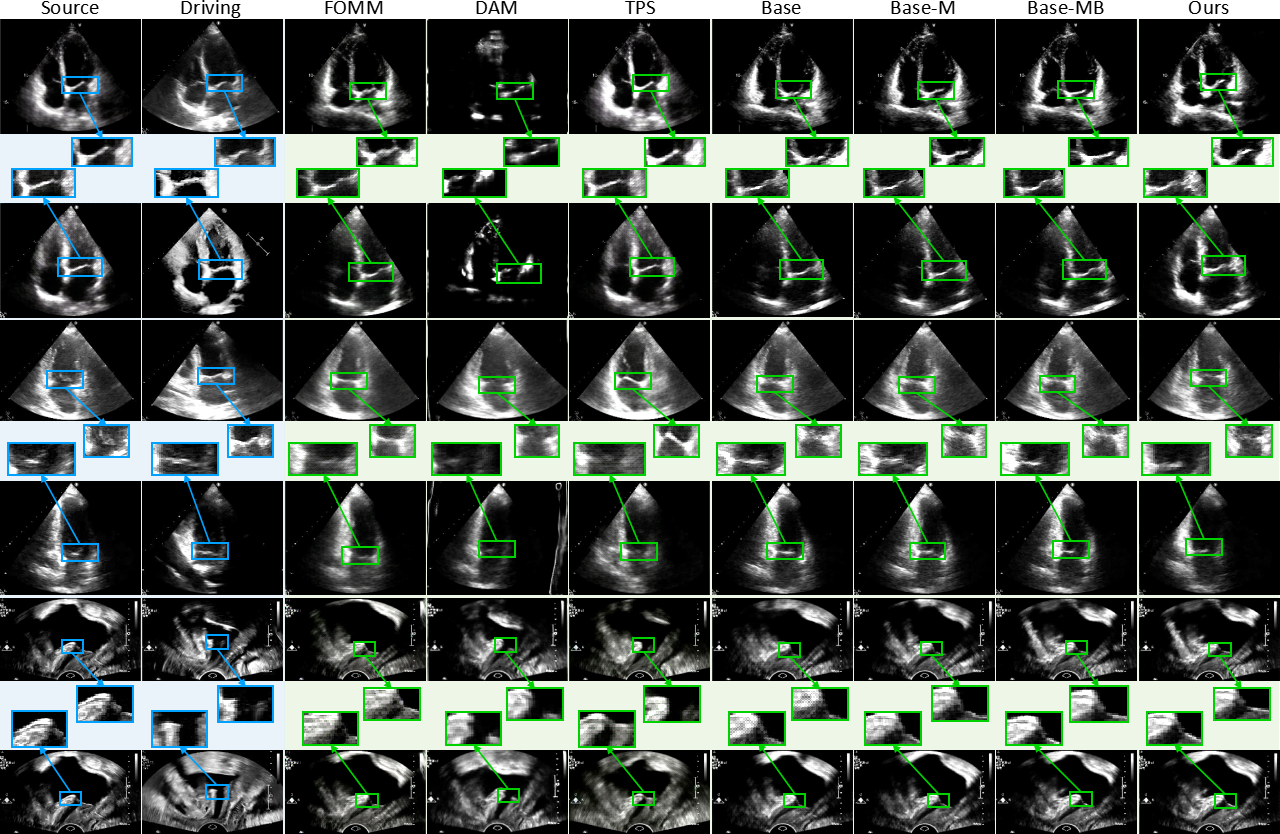}
    \caption{Qualitative comparison on animation task. From top to bottom, A4C heart echocardiography, A2C heart echocardiography, and pelvic floor.}
    \label{fig:prediction_result}
\end{figure*}
\textbf{Comparison on Generating Short US Videos.}
Online learning operates at the video level and requires high memory consumption. To avoid memory exhaustion, we divide US videos longer than 16 frames into multiple non-overlapping 16-frame videos, which serve as sub-driving videos for synthesizing US videos during testing. Each method synthesized 440 A4C, 321 A2C, and 202 pelvic floor US videos for the animation task, and 542 A4C, 198 A2C, and 126 pelvic floor US videos for the reconstruction task. Table~\ref{tab:COMPARISION} shows the quantitative results for both the reconstruction and animation tasks in the generation of 16-frame videos. The results of \ourmodel~reflect the addition of online learning to Base-MB, which achieves the best FVD on all three datasets, suggesting that \ourmodel~can generate videos with superior temporal continuity. However, the FID of \ourmodel~was slightly unsatisfactory. We assume that this can be attributed to the significant adjustments made to the source images to improve animation in the driving video. These adjustments complexify the image synthesis process, ultimately diminishing the realism of synthesized images. \par
Figs.~\ref{fig:reconstraction_result} and~\ref{fig:reconstraction_fluid} represent qualitative results for the reconstruction task. Fig.~\ref{fig:prediction_result} presents qualitative results for the animation task. 
\ourmodel~performed well on all three datasets with almost no artifacts (e.g., checkerboard effect \cite{deconvolution}), whereas FOMM, DAM, and TPS all exhibited artifacts (Fig.~\ref{fig:prediction_result}). Furthermore, \ourmodel~approximately preserved the anatomy, with the results closely resembling the GT (Fig.~\ref{fig:reconstraction_result} and Fig.~\ref{fig:reconstraction_fluid}). Although DAM synthesized A2C echocardiography and pelvic floor videos rather nicely, it failed to reconstruct or animate A4C echocardiography videos. We speculate that this can be attributed to the presence of excessive black areas in the A4C videos, resulting in the failure of learning useful image information. Additionally, as observed from Fig.~\ref{fig:reconstraction_fluid}, the video predicted by \ourmodel~is relatively fluid, which further indicates that \ourmodel~generates videos with better temporal continuity.\par
\begin{table}[!htbp]
\centering
\caption{Comparison with the state-of-the-art on animating long video.}
\setlength{\tabcolsep}{3.2mm}{
    \begin{tabular}{ccc|cc}
        \toprule  
        \multirow{2}{*}{Method}& \multicolumn{2}{c}{A4C}  & \multicolumn{2}{c}{A2C}\\
           \cline{2-5}  
           &FVD$\downarrow$&FID$\downarrow$&FVD$\downarrow$ &FID$\downarrow$ \\
           \hline
            FOMM & {\color{blue} 975.62} & 247.42 & 2027.87 & 250.51 \\
            TPS & 1031.74 & 205.41 & 2208.71 & 196.06 \\
            DAM & 1711.30 & 281.87 & 2188.96 & 273.49 \\
            Base-MB & 1004.99 & {\color{blue} 176.87} & {\color{blue} 1946.17} & {\color{blue} 195.49} \\
           \bottomrule 
     \end{tabular}}
 \label{tab:long_video_result}
\end{table}
\textbf{Comparison on Generating Long US Videos.}
The disparity in anatomic structure between the source and driving tends to increase in longer videos, posing a greater challenge in the generation of realistic US videos. To assess the effectiveness of our approach in generating long videos, we conducted an analysis utilizing Base-MB owing to memory constraints.\par
In addition to the US video used for model training and testing, we collected eight videos from four mitral regurgitation cases (each case contained one A2C and one A4C heart echocardiography video). We used these videos as driving and randomly selected US video frames from the testing set as source images. This process generated 46 A4C and 27 A2C heart echocardiographic videos under each synthesis method. The frame range for the synthesized A4C videos was 82-343, whereas that for the synthesized A2C videos was 86-192. The results summarized in Table~\ref{tab:long_video_result} indicate that the videos generated by Base-MB are more realistic. In addition, the temporal coherence of those videos is comparable to that of videos generated by FOMM \cite{FOMM} in the A4C dataset.\par
\textbf{User Study.}
\begin{table*}[!htbp]
\centering
\caption{Quantitative results of user study. Doc is short for Doctor. Fidelity reflects the overall realism of generated US videos, while fluidity represents the smoothness of motion within the anatomic structure, and texture refers to the realism of textures.}
    \begin{tabular}{c|c|cccccccccc|c}
        \toprule 
           &Dataset & Method &  Doc 1  &  Doc 2 &  Doc 3  &  Doc 4 & Doc 5 &Doc 6 &Doc 7 &Doc 8 &Doc 9 & Average \\
           \cline{1-13} 
           \multirow{12}{*}{Fidelity} & \multirow{4}{*}{A4C} 
             & GT & 5.00 & 4.93 & 4.11 & 3.59 & 4.41 & 4.22 & 4.30 & 4.98 & 4.20 & 4.42\\ 
             & & FOMM & 3.00 & 3.33 & 2.65 & 1.96 & 3.43 & {\color{blue}1.57} & 2.39 & 1.35 & 1.50 & 2.35\\ 
           & & TPS  & 2.76 & 3.00 & 2.17 & 1.65 & 2.59 & 1.00 & 2.04 & 2.26 & 2.17 & 2.18\\ 
            & & \ourmodel & {\color{blue}4.50} & {\color{blue}3.85} & {\color{blue}4.07} & {\color{blue}2.83} & {\color{blue}4.07} & 1.50 & {\color{blue}4.11} & {\color{blue}3.89} & {\color{blue}3.46} & {\color{blue}3.58}\\
            \cline{2-13}
             &\multirow{4}{*}{A2C} 
           & GT & 4.53 & 4.92 & 3.97 & 3.33 & 4.86 & 4.53 & 4.31 & 3.75 & 4.42 & 4.29\\
           && FOMM & 3.00 & 2.42 & 1.61 & 1.25 & 3.00 & 1.25 & 1.33 & 2.11 & 1.44 & 2.13\\
          && TPS  & 2.72 & 1.69 & 1.25 & 1.36 & 2.06 & 1.19 & 1.83 & 1.47 & 2.56 & 1.79 \\ 
        & & \ourmodel & {\color{blue}3.56} & {\color{blue}4.08} & {\color{blue}3.92} & {\color{blue}2.44} & {\color{blue}3.81} & {\color{blue}1.67} & {\color{blue}3.75} & {\color{blue}3.72} & {\color{blue}3.92} & {\color{blue}3.43} \\
            \cline{2-13} 
                &\multirow{4}{*}{Pelvic} 
                & GT & 4.92 & 4.17 & 4.5 & 4.58 & 4.00 & 4.75 & 4..83 & 4.75 & 4.92 & 4.60\\
                 && FOMM & 3.00 & 3.32 & 1.95 & 3.32 & 1.16 & 1.84 & 2.53 & 2.42 & 1.79 & 2.37\\
              & & TPS & 2.74 & 3.11 & 3.00 & 3.84 & 2.11 & 2.79 & 3.05 & 2.74 & 2.26 & 2.85 \\ 
              & & \ourmodel& \color{blue}4.26 & {\color{blue}3.68} & {\color{blue}3.21} & {\color{blue}4.00} & {\color{blue}3.00} & {\color{blue}4.53} & {\color{blue}3.89} & {\color{blue}4.26} & {\color{blue}4.21} & {\color{blue}3.89}\\ 
            \cline{1-13}  
            \multirow{12}{*}{Fluidity} & \multirow{4}{*}{A4C} 
             & GT & 4.98 & 4.93 & 4.26 & 2.91 & 4.41 & 4.15 & 4.41 & 4.98 & 4.20 & 4.36 \\ 
             &  & FOMM & 2.96 & 3.11 & 2.85 & 1.41 & 3.26 & {\color{blue}1.78} & 2.78 & 1.48 & 1.89 & 2.39 \\ 
           &  & TPS  & 2.50 & 2.30 & 2.43 & 1.22 & 2.43 & 1.07 & 2.50 & 2.30 & 2.11 & 2.10 \\ 
           & & \ourmodel &  {\color{blue}4.07} &  {\color{blue}3.59} &  {\color{blue}4.07} &  {\color{blue}2.11} &  {\color{blue}3.98} & 1.41 & {\color{blue}4.30} & {\color{blue}3.70} & {\color{blue}3.41} & {\color{blue}3.40}\\
            \cline{2-13}  
            & \multirow{4}{*}{A2C} 
           & GT & 4.50 & 4.94 & 4.11 & 3.44 & 4.86 & 4.50 & 4.31 & 4.08 & 4.58 & 4.37\\
           & & FOMM & 2.33 & 2.33 & 1.44 & 1.39 & 2.83 & 1.38 & 1.48 & 2.39 & 1.83 & 2.10 \\
          & & TPS & 2.53 & 1.81 & 1.42 & 1.14 & 2.11 & 1.25 & 1.86 & 1.72 & 2.89 & 1.86\\ 
           & & \ourmodel & \color{blue}3.44 & {\color{blue}3.67} & {\color{blue}3.83} & {\color{blue}2.19} & {\color{blue}3.69} & {\color{blue}1.50} & {\color{blue}3.44} & {\color{blue}4.00} & {\color{blue}4.06} & {\color{blue}3.31} \\
            \cline{2-13}
                &\multirow{4}{*}{Pelvic} 
                & GT & 4.92 & 3.92 & 4.92 & 4.50 & 3.58 & 4.75 & 4.75 & 4.67 & 4.75 & 4.53 \\
                 & & FOMM & 2.68 & 3.32 & 1.63 & 3.16 & 1.47 & 2.16 & 3.05 & 2.58 & 1.95 & 2.44\\
               & & TPS & 2.84 & 3.05 & 4.00 & 3.95 & 2.37 & 3.05 & 2.74 & 2.79 & 2.42 & 3.08 \\ 
              & & \ourmodel & {\color{blue}4.21} & {\color{blue}3.42} & {\color{blue}4.21} & {\color{blue}4.00} & {\color{blue}2.84} & {\color{blue}4.53} & {\color{blue}4.26} & {\color{blue}4.21} & {\color{blue}4.05} & {\color{blue}3.98}\\ 
              \cline{1-13}
            \multirow{12}{*}{Texture} & \multirow{4}{*}{A4C} 
             & GT & 5.00 & 4.91 & 3.93 & 2.87 & 4.41 & 4.20 & 4.15 & 4.98 & 4.17& 4.29 \\ 
             &  & FOMM & 2.43 & 2.61 & 2.22 & 1.35 & 2.70 & 1.26 & 2.13 & 1.33 & 1.30 & 1.93\\ 
           &  & TPS  & 1.48 & 2.00 & 1.87 & 1.22 & 2.28 & 1.00 & 1.78 & 1.74 & 2.02 & 1.71\\ 
           & & \ourmodel & {\color{blue}4.85} & {\color{blue}3.24} & {\color{blue}3.76} & {\color{blue}2.20} & {\color{blue}4.13} & {\color{blue}1.74} & {\color{blue}3.91} &{\color{blue}4.20} & {\color{blue}3.54} & {\color{blue}3.51}\\
            \cline{2-13}
            & \multirow{4}{*}{A2C} 
           & GT & 4.39 & 4.83 & 3.67 & 3.31 & 4.92 & 4.69 & 4.25 & 3.56 & 4.53 & 4.24 \\
           & & FOMM & 2.78 & 1.94 & 1.25 & 1.44 & 2.25 & 1.44 & 1.08 & 1.48 & 1.28 & 1.72\\
          & & TPS & 2.00 & 1.19 & 1.17 & 1.14 & 1.59 & 1.11 & 1.56 & 1.28 & 2.56 & 1.51 \\ 
           & & \ourmodel & {\color{blue}2.92} & {\color{blue}4.64} & {\color{blue}3.42} & {\color{blue}2.25} & {\color{blue}4.22} & {\color{blue}1.72} & {\color{blue}3.97} & {\color{blue}3.56} & {\color{blue}3.94} & {\color{blue}3.40}\\
            \cline{2-13} 
                &\multirow{4}{*}{Pelvic} 
                & GT & 4.92 & 4.00 & 4.50 & 4.50 & 3.83 & 4.58 & 4.83 & 4.50 & 4.92 & 4.51\\
                 & & FOMM & 2.16 & 2.21 & 1.42 & 2.63 & 1.11 & 1.84 & 2.21 & 1.68 & 1.53 & 1.87\\
               & & TPS & 1.42 & 2.32 & 2.95 & 3.84 & 1.84 & 2.32 & 2.26 & 2.32 & 1.37 & 2.29\\ 
              & & \ourmodel & {\color{blue}4.00} & {\color{blue}3.26} & {\color{blue}3.16} & {\color{blue}4.05} & {\color{blue}2.68} & {\color{blue}4.26} & {\color{blue}3.74} & {\color{blue}3.84} & {\color{blue}4.32} & {\color{blue}3.70}\\ 
           \bottomrule 
     \end{tabular}
 \label{tab_user_study}
\end{table*}
\begin{figure*}[!t]
    \centering
    \includegraphics[width=.85\linewidth]{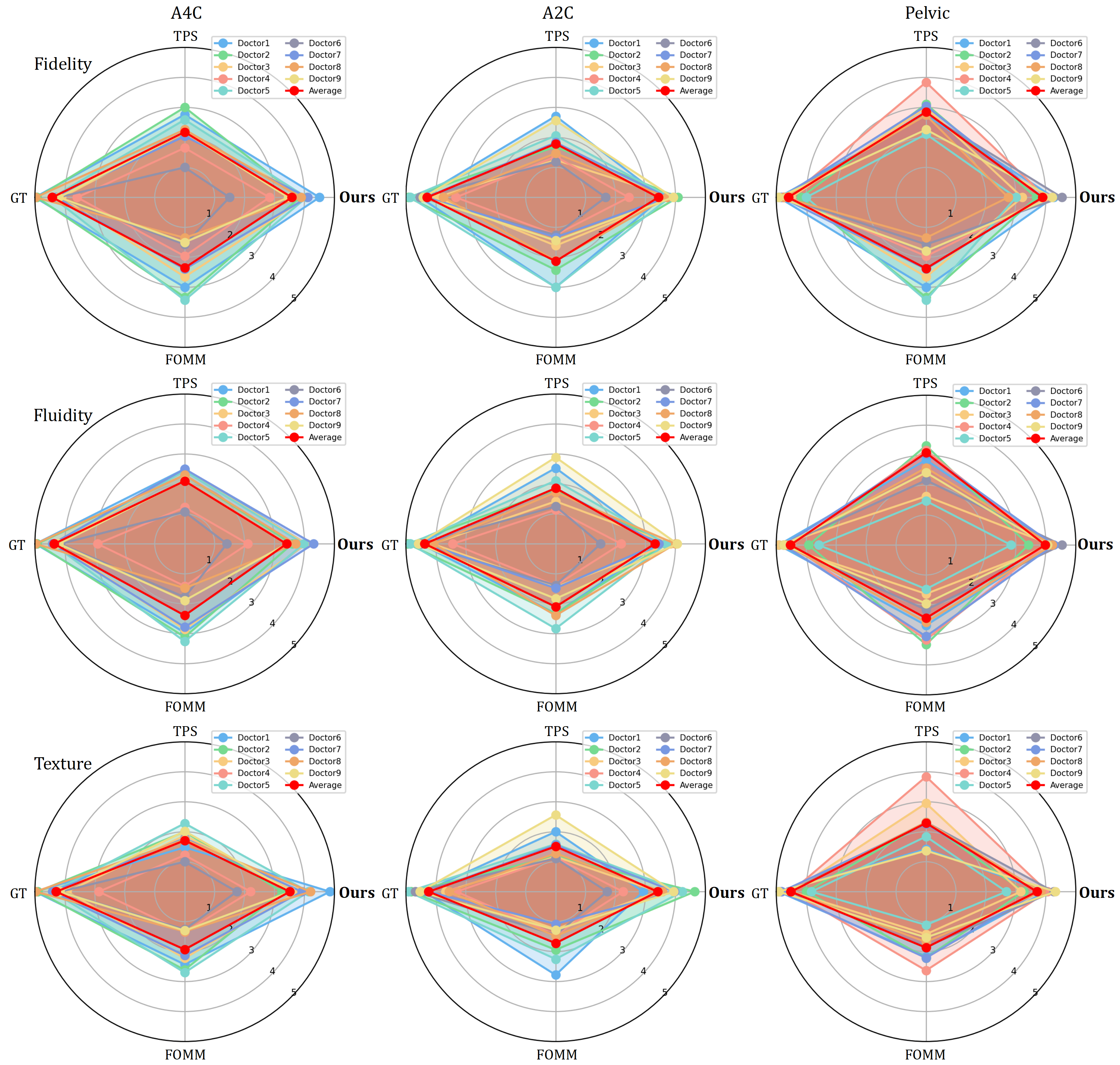}
    \caption{Radar charts showing the results of the user study. The scores of \ourmodel~ are the closest to GT compared to other competitors.}
    \label{fig:User_Study_fig}
\end{figure*}
\begin{table*}[!htbp]
\centering
\caption{US video classification Results. ACC, Pre, F1, Sen, and Spe denote accuracy, precision, F1-score, sensitivity, and specificity respectively.}
    \begin{tabular}{c|c|c|cccc|cccc}
        \toprule 
        \multirow{2}{*}{Model} & \multirow{2}{*}{Method} & \multirow{2}{*}{ACC (\%)} & \multicolumn{4}{c}{A4C} & \multicolumn{4}{c}{A2C}\\
        \cline{4-11} 
        & &  &Pre (\%) &F1 (\%) & Sen (\%) &Spe (\%)  &Pre (\%) & F1 (\%) & Sen (\%) &Spe (\%) \\
           \hline
           \multirow{4}{*}{ R(2+1)D }
           &FOMM & 89.34 & 86.11 & 90.51 & {\color{blue}95.38} & 82.46 & {\color{blue}94.00} & 87.85 & 82.46 &  {\color{blue}95.38} \\
            &TPS & 80.33 & 80.60 & 80.00 & 83.08 & 77.19 & 80.00 & 78.57 & 77.19 & 83.08 \\
            &DAM & 62.30 & 78.79 & 53.06 & 40.00 & 87.72 & 56.18 & 68.49 & 87.72 & 40.00 \\
            &\ourmodel & {\color{blue}91.80} & {\color{blue}92.31} & {\color{blue}92.31} & 92.31 & {\color{blue}91.23} & 91.23 & {\color{blue}91.23} &{\color{blue}91.23} & 92.31 \\
            \hline
            \multirow{4}{*}{ SlowFast }
           &FOMM & 67.21 & {\color{blue}100.00} & 55.56 & 38.46 & {\color{blue}100.00} & 58.76 & 74.03 & {\color{blue}100.00} & 38.46\\
            &TPS  & 46.72 & 50.00 & 5.80 & 3.08 & 96.49 & 46.61 & 62.86 & 96.49 & 3.08\\
            &DAM & 63.94 & 72.34 & 60.71 & {\color{blue}52.31} & 77.19  & 58.67 & 66.67 & 77.19 & {\color{blue}52.31} \\
            &\ourmodel & {\color{blue}70.49} & 93.94 & {\color{blue}63.27} & 47.69  & 96.49 & {\color{blue}61.80} & {\color{blue}75.34} & 96.49 & 47.69 \\
             \hline
            \multirow{4}{*}{ CSG-3DCT }
           &FOMM & 93.44 & 95.24 & 93.75 & 92.31 & 94.74 & 91.53 & 93.10 & 94.74 & 92.31\\
            &TPS  & 86.88  & 85.51 & 88.06 & 90.77 & 82.46 & 88.68 & 85.45 & 82.46 & 90.77\\
            &DAM & 47.54 & {\color{blue}100.00} & 3.03 & 1.54 & {\color{blue}100.00} & 47.11 & 64.04 & {\color{blue}100.00}& 1.54\\
            &\ourmodel & {\color{blue}95.08} & 94.03 & {\color{blue}95.45} & {\color{blue}96.92} & 92.98 & {\color{blue}96.36} & {\color{blue}94.64} & 92.98 & {\color{blue}92.92} \\
           \bottomrule
     \end{tabular}
 \label{tab:us_video_class}
\end{table*}
The conventional indices used to evaluate generated US video quality in our study cannot fully capture the intricacies of human perception. To obtain a more comprehensive fidelity assessment, we conducted a user study. This approach allowed us to obtain valuable insights and subjective evaluations from human observers, thereby complementing the objective metrics. Among all comparative experiments, TPS \cite{TPS} and FOMM \cite{FOMM} performed relatively well; therefore we chose them for the comparison.\par
We synthesized 46 A4C and 36 A2C heart echocardiographic videos, along with 19 pelvic floor US videos using each method. A total of 184 A4C and 144 A2C heart echocardiographic and 76 pelvic floor US videos were considered in the user study. We invited nine experienced physicians to rate each video on a five-point scale in terms of fidelity, fluidity, and texture, with higher scores indicating higher realism. \par
\begin{figure}[!t]
    \centering
    \includegraphics[width=1\linewidth]{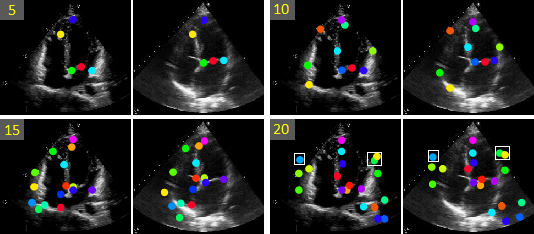}
    \caption{Comparison of keypoint estimation by Base-MB, with the number of keypoints alternating between 5, 10, 15, and 20.}
    \label{fig:heart_kp_visual}
\end{figure}
The results of the user study are presented in Table~\ref{tab_user_study}, which shows that the average scores of \ourmodel~on all datasets in terms of fidelity, fluidity, and texture were the highest. The corresponding radar plots are presented in Figs.~\ref{fig:User_Study_fig}. It is apparent that \ourmodel~exhibits superior performance compared to TPS and FOMM in the A4C and A2C datasets. However, in the pelvic floor dataset, the difference in performance between \ourmodel~and TPS was relatively small. We assume that this can be attributed to the greater range and magnitude of motion exhibited by the heart compared to those of the pelvic floor. The synthesis of A4C and A2C US videos is inherently more challenging and therefore more effective for evaluation. From these results, we can conclude that videos synthesized using \ourmodel~are perceived as more realistic than those synthesized using other methods from a human perspective. \par
\begin{figure*}[!t]
    \centering
    \includegraphics[width=0.9\linewidth]{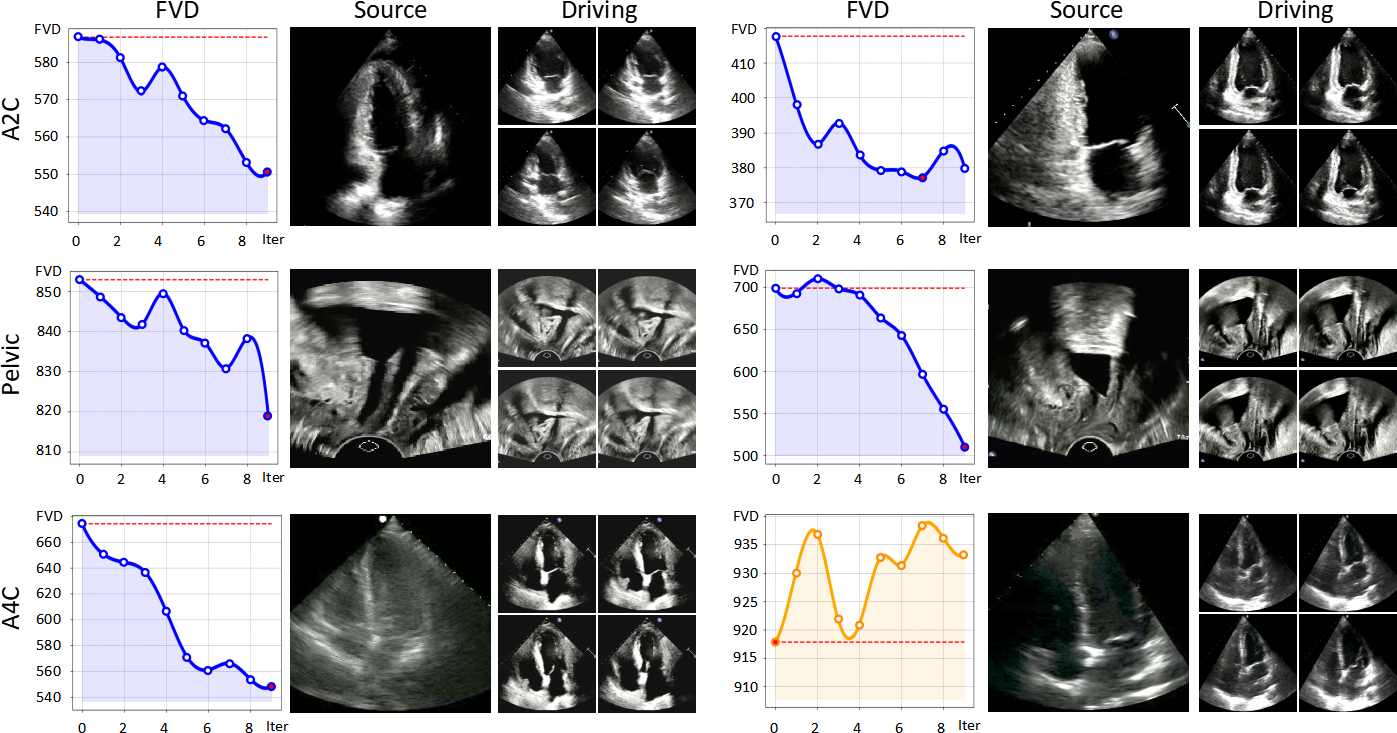}
    \caption{Changes in FVD values of generated videos during the online learning process. The red line in the chart represents the FVD value of the video generated prior to online learning. Red dots indicate the final FVD. The corresponding driving frames and source images of each folding line chart are shown on its right side. Good examples are shown in blue, whereas poor example is shown in orange. Iter is short for iteration.}
    \label{fig:ONLINE_result}
\end{figure*}
\textbf{Domain Differences Between Real and Generated Videos.} 
The performance of a model on a testing set can be affected by domain differences between the training (source domain) and the testing (target domain) sets \cite{dorent2023crossmoda}. A smaller gap between the source and target domains generally leads to better performance. To further compare the domain differences between videos synthesized by different methods and real videos, we conducted A4C and A2C US video classification experiments. Specifically, we synthesized 216 A2C and 272 A4C heart echocardiographic US videos using each distinct synthesis method to construct training sets. The testing set comprised 65 A4C and 57 A2C heart real US videos.\par
During the training and testing phases, each video was sampled eight frames uniformly. To ensure a fair comparison, the only variable used in the classification experiment was the training set. All classification models were trained for 30 epochs, and the weight of the last epoch was used for testing.
The results of the US video classification experiments are listed in Table~\ref{tab:us_video_class}. R(2+1)D \cite{R3D} and SlowFast \cite{slowfast} are classic video classification models, whereas CSG-3DCT \cite{XINRUI} was designed specifically for US video classification. It is evident from Table~\ref{tab:us_video_class} that \ourmodel~achieved the highest accuracy among the compared methods. This indicates that \ourmodel~can generate US videos closer to the source domain.
\subsection{Analysis of Impact of Keypoint Number}
The number of keypoints represents an important hyperparameter in our study. To analyze the impact of this parameter on overall performance in the training process, we conducted experiments with varying keypoint numbers (5, 10, 15, and 20) while keeping all other parameters constant. All experiments were conducted on the A4C heart video dataset using Base-MB. As shown in Table~\ref{Tab:heart-4c_alb_study}, overall performance initially improves with an increase in the number of keypoints, before declining, when said number exceeds a certain threshold. Furthermore, an increase in the number of keypoints results in a decrease in the fidelity of frames in the generated videos.\par
To analyze these results, we visualized keypoints predicted by the keypoint detector as shown in Fig.~\ref{fig:heart_kp_visual}. As the number of keypoints increases, the coverage area expands. A wider coverage area outside the black BG area can enhance the accuracy in estimating the motion field. However, when the number of keypoints reaches 20, a portion of them appear within the black BG area (indicated by the white boxes). The presence of keypoints in the black BG area can negatively affect the evaluation of a dense motion field. Furthermore, increasing the number of keypoints leads to more complex affine transformations, thereby increasing the potential for pixel position alterations between the source and generated frames, adversely affecting the realism of generated frames.\par
\begin{table}[!htbp]
\centering
\caption{Video reconstruction and animation results for different numbers of keypoints on the A4C echocardiography dataset. KP denotes the number of keypoints.}
    \small 
    \begin{tabular}{c|l|cccc}
        \toprule 
          \multicolumn{2}{c}{KP}& 5& 10 & 15 & 20\\
            \cline{1-6}
           \multirow{6}{*}{Reconstruction} &FVD$\downarrow$ & 253.81& 243.73& {\color{blue}228.75}& 238.18 \\
           &FID$\downarrow$ & {\color{blue}28.28}& 30.79 & 30.83 & 30.84\\
           &LPIPS$\downarrow$  & 0.0328 & 0.0326 & {\color{blue}0.0319}& 0.0323\\
           &L1$\downarrow$ & 0.0192& 0.0187 & {\color{blue}0.0186}& 0.0187 \\
           &PSNR$\uparrow$ & 27.00 & 27.23 & {\color{blue}27.33} & 27.29\\
           &SSIM$\uparrow$ & 0.8362& 0.8376& {\color{blue}0.8394} & 0.8356\\
           \cline{1-6}
           \multirow{2}{*}{Animation} &FVD$\downarrow$  & 427.64 & 421.41& {\color{blue}407.19} & 418.57\\
           &FID$\downarrow$ & {\color{blue}76.21}& 76.67 
           & 76.74& 77.05 \\
           \bottomrule 
     \end{tabular}
 \label{Tab:heart-4c_alb_study}
\end{table}
\subsection{Analysis of Keypoint Motion Trajectory Constraints}
To gain insight into the impact of motion trajectory constraints, we monitored the FVD for several generated videos during the online learning process. The results are presented in Fig.~\ref{fig:ONLINE_result}. During the online learning process, the motion trajectories of keypoints in the generated video were gradually aligned with those of keypoints in the driving video. Consequently, most generated videos exhibited enhanced fluency compared to the original generated videos. However, there were instances wherein the FVD of certain generated videos decreased, as seen in Fig.~\ref{fig:ONLINE_result}. This can be attributed to significant variations in size between anatomic structures in the source image and driving video. In this case, if the motion trajectories of the keypoints in the driving video are simply employed to constrain those of keypoints in the generated video, the latter may be lead to unreasonable. The decrease in FVD of the generated videos may also be attributed to the unclear boundaries of anatomic structures. In such cases, the accurate localization of keypoints becomes more challenging. Applying constraints to the keypoint motion trajectory under such circumstances can further exacerbate the instability of keypoint prediction.
\subsection{Analysis of Time Efficiency} 
To gain insights into the time efficiency of \ourmodel, we calculated the average time taken by Base, Base-M, Base-MB, and \ourmodel~to synthesize a 10-frame A4C heart US videos. The keypoint number is set to 15 for all methods. The results, as shown in Table~\ref{tab:time_consuming}, demonstrate that \ourmodel~can efficiently synthesize a 10-frame US video in less than 1 second. The inclusion of a multi-scale discriminator increases the synthesis time by 20.85 $ms$, while the BG motion predictor and online learning process lead to an increase of 382.60 $ms$ and 156.54 $ms$, respectively. This indicates that \ourmodel~effectively improves the fluidity and realism of generated US videos at the cost of only a minor increase in time.\par
\begin{table}[!htbp]
\centering
\caption{The mean(std) time of different methods on synthesizing a 10-frame A4C US video.}
\setlength{\tabcolsep}{1.4mm}{
    \begin{tabular}{c|cccc}
        \toprule  
         & Base   & Base-M & Base-MB & \ourmodel\\
           \cline{1-5} 
           Time/$ms$ & $224.78_{22.95}$ & $245.63_{33.22}$ & $628.23_{44.97}$ & $784.77_{28.84}$\\
           \bottomrule 
     \end{tabular}}
 \label{tab:time_consuming}
\end{table}
\section{Discussion}
Though \ourmodel~can synthesize US videos with high fidelity, there are certain rooms for continuous improvement.\par
\textbf{Video Resolution.} The resolution of US videos synthesized by \ourmodel~was $256\times256$, which is smaller than the resolution typically used in clinical practice. When we attempted to generate A4C echocardiographic videos with a resolution of 512 pixels, we observed distortions in the synthesis. This issue may stem from the discriminator effortlessly distinguishing between real and synthesized images, consequently hindering the overall performance of the generator.\par
\textbf{Factors Impacting the Synthesis Quality.} A slight degradation in the synthesis quality may be observed when the source and driving are derived from different machine models or hospitals. Besides, despite our efforts in constraining the motion trajectories of anatomic structures, certain synthesized videos may exhibited implausible movement.\par
\textbf{Selection of Image Synthesis Method.} The diffusion model represents an alternate approach for generating US videos \cite{molad2023dreamixdiffusion}. By defining a Markov chain of diffusion steps to slowly add noise to the data and learn to reverse the diffusion process, diffusion models generate samples with high levels of quality and diversity \cite{SurveyDiffusion}. But diffusion model might not suit the dual-decoder-based generation of US images and lead to the loss of important textural features. Furthermore, diffusion models are associated with high computational burdens and low speeds. Although video synthesis does not necessarily require a real-time approach, current video synthesis frameworks are complex and computationally intensive enough. The introduction of a diffusion model for video synthesis may further amplify the demand for computational resources.\par
\section{Conclusion}
US is widely employed as a medical imaging modality owing to its real-time capabilities, safety, and cost efficiency. However, it is challenging for sonographers to extract diagnostic information from the morphological and kinematic characteristics of anatomic structures using US videos. Moreover, the scarcity of such videos may hinder the effective teaching of detection skills during sonographer training, potentially impacting the diagnosis of diseases. To solve this problem, we propose a novel online feature-decoupling framework called \ourmodel~for high-fidelity US video synthesis with the desire to enhance medical US training and disease modeling. \par
In this study, we introduced anatomic structure information into keypoint learning in a weakly-supervised manner. It was for better preserving anatomic integrity and mimicking the anatomic motion in the driving video more accurately within the minimal labeling burden. In addition, a dual-decoder approach was devised to better preserve the content and textural information in generated US frames. We then proposed an adversarial training strategy with a multi-scale discriminator to enhance the sharpness and fine details of the synthesized videos. Finally, we introduced an innovative online learning strategy to enhance the fluency of the generated videos. Through validation and user studies conducted on in-house echocardiographic and female pelvic floor US videos, \ourmodel~demonstrated its capability to synthesize US videos with minimal artifacts and a high level of fidelity.\par
In the future, we aim to address the limitations of our study and attempt to perform a more controlled synthesis of US videos to further meet the demands of clinical US training. For example, inspired by \cite{LUO2022102335}, we will try to introduce spectral regularization to increase the quality of synthesized videos.

\bibliographystyle{IEEEtran}

\bibliography{IEEEabrv}{}

\end{document}